\documentclass[preprint,aps,amsmath,amssymb,nofootinbib]{revtex4}
\usepackage[mathscr]{euscript}
\usepackage{threeparttable}
\usepackage{booktabs}
\usepackage{subfigure}
\usepackage{bm}
\usepackage{textcomp}
\usepackage{graphicx}
\usepackage{overpic}
\usepackage{multirow}
\usepackage{floatrow}
\usepackage{float} 
\usepackage{amsmath,amssymb,amsthm}
\usepackage{cases}
\usepackage{xcolor}
\usepackage[colorlinks=true,linkcolor=red]{hyperref}
\usepackage[normalem]{ulem}
\newfloatcommand{capbtabbox}{table}[][\FBwidth]

\begin{document}

\title{Phantom-Divide Crossing in Exponentially Coupled Quintessence and the Role of Neutrino-Mass Freedom} 

\author{Jincheng Wang\footnote{J.C.Wang@hunnu.edu.cn}}
	
\affiliation{Department of Physics, Key Laboratory of Low Dimensional Quantum Structures and Quantum Control of Ministry of Education, and Hunan Research Center of the Basic Discipline for Quantum Effects and Quantum Technologies, Hunan Normal University, Changsha, Hunan 410081, China} 

\author{Hongwei Yu\footnote{Corresponding author: hwyu@hunnu.edu.cn}}
\affiliation{Department of Physics, Key Laboratory of Low Dimensional Quantum Structures and Quantum Control of Ministry of Education, and Hunan Research Center of the Basic Discipline for Quantum Effects and Quantum Technologies, Hunan Normal University, Changsha, Hunan 410081, China}

 \author{Puxun Wu\footnote{Corresponding author: pxwu@hunnu.edu.cn} }
\affiliation{Department of Physics, Key Laboratory of Low Dimensional Quantum Structures and Quantum Control of Ministry of Education, and Hunan Research Center of the Basic Discipline for Quantum Effects and Quantum Technologies, Hunan Normal University, Changsha, Hunan 410081, China}
	
\begin{abstract}
We investigate a quintessence dark-energy model with an exponential potential and an exponential coupling to cold dark matter (CDM), hereafter referred to as the CQ-EXP model,  using  Planck CMB, DESI BAO, and DES-Dovekie supernova observations.  We also examine  how variations in the neutrino mass sector affect the constraints. 
When the neutrino mass sum is fixed at $\sum m_\nu=0.06$ eV, the data favor a coupling between quintessence and CDM, with the coupling parameter $\beta$ deviating from zero at more than $3\sigma$.
 In particular,  the observations favor the $\beta<0$ branch, where the energy transfer between the two dark sectors changes sign and the effective equation of state (EoS) of dark energy crosses the phantom divide, $w=-1$. When the effective  neutrino mass parameter $\sum m_{\nu,\mathrm{eff}}$  is treated as a free parameter, the data show a preference for  negative values of $\sum m_{\nu,\mathrm{eff}}$. This additional freedom weakens the preference for the coupling between quintessence and CDM and leads to nearly identical values of  $\chi^2_{\rm min}$ for the CQ-EXP models with $\beta>0$ and $\beta<0$,  corresponding respectively to models without and with  phantom-divide crossing in the effective EoS.  Both values are slightly   larger than that obtained in  the $w_0w_a$CDM model, indicating that the CQ-EXP model cannot be statistically distinguished from the $w_0w_a$CDM model with the data considered here.  Therefore, when $\sum m_\nu$ is fixed, current observations favor the CQ-EXP model with phantom-divide crossing. In contrast,  when negative values of  $\sum m_{\nu,\mathrm{eff}}$ are allowed,  a CQ-EXP dark energy without crossing $w=-1$  can also provide an effective explanation of the latest observations.
 \end{abstract}

	
	\maketitle
\section{Introduction}
\label{sec1}

Recent measurements of baryon acoustic oscillation (BAO) from the Dark Energy Spectroscopic Instrument (DESI)~\cite{Adame2024desi}, particularly those from Data Release 2 (DR2)~\cite{DESIDR2}, have provided  compelling evidence in favor of dynamical dark energy rather than a simple cosmological constant $\Lambda$.  This evidence reaches a significance of  about $3.1\sigma$~\cite{DESIDR2} when  the BAO data from DESI DR2 are combined with  Planck cosmic microwave background (CMB) data  to constrain the $w_0w_a$CDM model.   This model  incorporates  cold dark matter (CDM) and dark energy described by the Chevallier-Polarski-Linder (CPL) parametrization for its equation of state (EoS): $w(a)=w_0+w_a z/(1+z)$~\cite{chevallier2001accelerating,linder2003exploring},      where $w_0$ and $w_a$ are constants and $z$ is the redshift. For  $w_0=-1$ and $w_a=0$, the $w_0w_a$CDM model reduces to the standard $\Lambda$CDM model.  With the inclusion of additional Type Ia supernovae (SNe Ia)  samples,  the significance of  dynamical dark energy  ranges from $2.8\sigma$ to $4.2\sigma$~\cite{DESIDR2}. The combination of DESI BAO, Planck CMB, and SNe Ia favors the constraints $w_0>-1$ and $w_0+w_a<-1$~\cite{DESIDR2b,gu2025desidr2dde,Wang2025,wang2026planckdesi}. This  indicates   that the EoS of dark energy evolves from below $-1$ at earlier times to above $-1$ at late times during cosmic expansion. 

Dynamical dark energy can be realized using   scalar fields, such as quintessence, which is described  by a canonical scalar field~\cite{ratra1988cosmological,Caldwell1998,Zlatev1998,copeland2006dynamics}. However, the EoS of a quintessence scalar field $\phi$ satisfies  $w_\phi\geq -1$~\cite{vikman2005can,caldwell2005crossing}, making it unable to explain scenarios where $w<-1$, as indicated by  DESI BAO observations.  One way to realize $w<-1$ is to introduce scalar fields with negative kinetic energy, known as phantom fields~\cite{caldwell2002phantom}.  However, negative kinetic energy generally leads to quantum instabilities~\cite{carroll2003phantom,cline2004phantom}. 

To avoid such instabilities while still accommodating the observational preference for phantom-like behavior, one can generalize quintessence by introducing an interaction between dark energy and dark matter~~\cite{wang2024review,amendola2000coupled,pettorino2008cqe,li2026strongevidence,samanta2025exploring,beltran2026ceq}.  In this framework, the effective EoS of dark energy can cross the phantom divide line, permitting  effective values below $-1$ without requiring a phantom scalar field~\cite{wang2005holographic,petri2025darkdegeneracy,chakraborty2025hint,chakraborty2024hint,silva2025newconstraints,gomezvalent2026desi,antusch2026guide}. 

Among quintessence scenarios,  scalar fields with an exponential potential  are especially well studied~\cite{amendola2000coupled,beltran2026ceq,wetterich1988cosmology,copeland1998exponential,bhattacharya2024curved,tocchini2002stationary,gumjudpai2005dynamics}.  The exponential  potential is given by
\begin{eqnarray}\label{eq1}
V(\phi)=V_0 e^{-\alpha\phi/M_{\rm Pl}}.
\end{eqnarray}
Here,  $V_0$ sets the scale of dark energy  and is determined by matching the current dark energy density,  $\alpha$ is a constant that characterizes the slope of  the potential and is set to be greater than zero without loss of generality,  and $M_{\rm Pl}$ is the Planck mass. 

When  quintessence  is coupled to   CDM,  a commonly used  coupling takes  the exponential  form   $e^{-\beta\phi/M_{\rm Pl}}$~\cite{amendola2000coupled,pettorino2008cqe,anchordoqui2026desiportal}. This leads to an energy transfer between the two dark sectors  of the form 
\begin{equation}
Q=\frac{\beta}{M_{\rm Pl}}\rho_c\dot\phi .
\end{equation}
Here, an overdot denotes a derivative with respect to cosmic time $t$, $\beta$ is the coupling parameter, and $\rho_c$ is  the CDM energy density.    In coupled quintessence with an exponential potential and an exponential coupling, hereafter referred to as the CQ-EXP model, the branch with $\beta>0$ is usually  considered~\cite{amendola2000coupled,pettorino2008cqe,baldi2012codecs,beltran2026ceq,liu2024emtransfer}.    In this work, we extend the analysis to include the branch with $\beta<0$.  We discover that,  in the  CQ-EXP model, the $\beta<0$  branch exhibits more intriguing dynamics  than the $\beta>0$ branch. Specifically, for $\beta<0$, the energy transfer $Q$ between dark energy and dark matter can change sign. This occurs because $\dot{\phi}$  changes sign:  the interaction first drives the scalar field to roll up its potential before it subsequently rolls down. This dynamics allows the effective EoS of dark energy to cross the phantom divide, $w=-1$.
 The mechanism responsible  for this sign-changing  interaction is similar to that found in coupled quintessence models with a supergravity-inspired potential~\cite{baldi2011bouncing,wangSUGRA}. However, it differs from the mechanism in  quintessence  models with a coupling of the  form  $1+\beta \phi^2$~\cite{wang2026signswitch}, where the sign change arises because $Q\propto \frac{\phi}{1+\beta \phi^2}$, resulting in a reversal of sign as $\phi$ passes through the zero. In this paper, we test both branches of the CQ-EXP model, $\beta$ ($\beta>0$ and $\beta<0$), using  Planck CMB measurements~\cite{planck2018cosmo,rosenberg2022pr4camspec,carron2022pr4lensing}, DESI DR2 BAO data~\cite{DESIDR2}, and DES-Dovekie SNe Ia data~\cite{desdovekie2025,dessn5yr2024}. 
 
When  CMB data are used to constrain the  cosmological parameters, the sum of neutrino masses is often fixed to a fiducial value, such as $\sum m_{\nu}=0.06$ eV~\cite{planck2018cosmo}.  If this assumption is relaxed while imposing the physical prior $\sum m_\nu\geq 0$, observations tend to favor a vanishing neutrino mass sum~\cite{DESIDR2nu,du2025impacts}. Moreover, if one takes an additional step to treat the effective neutrino mass parameter  $\sum m_{\nu,\rm{eff}}$ as a free parameter and allow it to take negative values,  cosmological observations show a preference for negative  $\sum m_{\nu,\rm{eff}}$~\cite{DESIDR2nu,jhaveri2025negative,elbers2025negative,kibris2026negative}.  Although such values are unphysical in the standard interpretation, they can have important phenomenological implications, leading to different observational preferences in the $\Lambda$CDM and $w_0w_a$CDM frameworks~\cite{du2025impacts,elbers2025negative,kibris2026negative,du2025positive}.  Therefore,  it is crucial  to investigate how the CQ-EXP model responds to  variations  in the neutrino mass sector. Motivated by this,  we   constrain the CQ-EXP model  both with a fixed value of 
 $\sum m_{\nu}$ and  with $\sum m_{\nu,\rm{eff}}$ treated as a free parameter.   
 
The remainder of the paper is organized as follows. In Sec. II and Sec. III, we introduce the CQ-EXP model and discuss the treatment of neutrino masses. Sec. IV, we describe the observational data used in our analysis. Sec. V presents the results and discussions. Finally, Sec. VI  summarizes our conclusions.

\section{Coupled  Quintessence with an Exponential Potential}%
\label{sec2}

We consider a spatially flat Friedmann-Lema\^{\i}tre-Robertson-Walker cosmology~\cite{friedmann1922,lemaitre1927,robertson1935,walker1937}, described by the metric: $ds^2=-dt^2+a^2(t) \delta_{ij}dx^idx^j$. Here, $a(t)$ is the cosmic scale factor.  We assume that  a quintessence  scalar field $\phi$ is conformally coupled only to CDM, while baryons and radiation remain minimally coupled to $\phi$. After the conformal transformation, the dark sector action can be written as~\cite{amendola2000coupled,pettorino2008cqe}
\begin{equation}
	\begin{aligned}
		S ={} \int d^4x \sqrt{-g}\left[
		-\frac{1}{2}g^{\mu\nu}\partial_{\mu}\phi\partial_{\nu}\phi
		-V(\phi)
		\right] 
		+S_c\!\left[ \mathcal{A}^2(\phi){g}_{\mu\nu};\psi_c\right],
	\end{aligned}
	\label{action_cq}
\end{equation}
where $g$ is the determinant of the background metric $g_{\mu\nu}$,   $\psi_c$ denotes the CDM degrees of freedom,   $V(\phi)$ is the potential of quintessence given in Eq.~(\ref{eq1}), and $\mathcal{A}(\phi)$ is the scaling factor that connects the background metric and the conformally transformed metric. Here, $\mathcal{A}(\phi)$ is assumed to be   a commonly used exponential  coupling~\cite{amendola2000coupled,pettorino2008cqe}
\begin{eqnarray}
\mathcal{A}(\phi)\propto \exp \left(-\beta \frac{\phi}{M_{\rm Pl}} \right). 
\end{eqnarray} 

From the action given in Eq.~\eqref{action_cq}, we can derive that the energy densities of  CDM  and the scalar field  obey the following equations:
\begin{equation}
	\dot{\rho}_c+3H\rho_c=-Q, \quad \dot{\rho}_\phi+3H(\rho_\phi+p_\phi)=Q
	\label{rho_c_eq}
\end{equation}
with
\begin{equation}
	Q \equiv -\frac{d\ln \mathcal{A}}{d\phi}\,\rho_c\dot{\phi} =\beta \rho_c\frac{\dot{\phi}}{M_{\rm Pl}}.
	\label{Q_definition}
\end{equation}
Here,  $H=\dot{a}/a$ is the Hubble parameter.  $Q>0$ indicates that there is an energy transfer from  CDM to the quintessence field. Conversely, $Q<0$ signifies an energy transfer from the quintessence field to  CDM.  The CDM conservation equation can be formally integrated out,  yielding
\begin{equation}
	\rho_c=\rho_{c0}a^{-3}
	\frac{\mathcal{A}(\phi)}{\mathcal{A}(\phi_0)},
	\label{rho_c_integrated}
\end{equation}
where the subscript $0$ denotes the present value.

 The action Eq.~(\ref{action_cq}) can also give    the equation of motion  for the scalar field 
\begin{equation}
	\ddot{\phi}=-3H\dot{\phi}-V_{,\phi}
	-\rho_c\frac{d\ln \mathcal{A}}{d\phi}=-3H\dot{\phi}-V_{,\phi}
	+\beta \frac{\rho_c}{M_{\rm Pl}}.
	\label{kg_eq}
\end{equation}
The three terms on the right-hand side of Eq.~(\ref{kg_eq}) can be regarded as the Hubble friction due to cosmic expansion, the conservative force derived from the  potential  of the scalar field, and the effective force induced by the interaction between the two dark sectors, respectively. The Hubble friction  serves to dampen the motion of the scalar field. However, it does not determine the direction of the field's movement. Instead, this direction is dictated  by the signs and relative magnitudes of the last two terms. Since $-V_{,\phi}=\frac{\alpha}{M_{\rm Pl}}V(\phi)>0$, we find that the different signs of $\beta$  correspond to two physically distinct dynamical branches. When $\beta>0$, both  the conservative  force and the effective force    act in the same direction within the field space. This alignment can enhance the  acceleration of  the scalar field, promoting movement in a direction that is supported by the potential. Conversely, when  $\beta<0$, the conservative and effective forces act in opposite directions. In this scenario,  the interaction can  drive the scalar field against the direction preferred by the underlying potential.

\begin{figure}[t]
\centering
\includegraphics[width=0.45\textwidth]{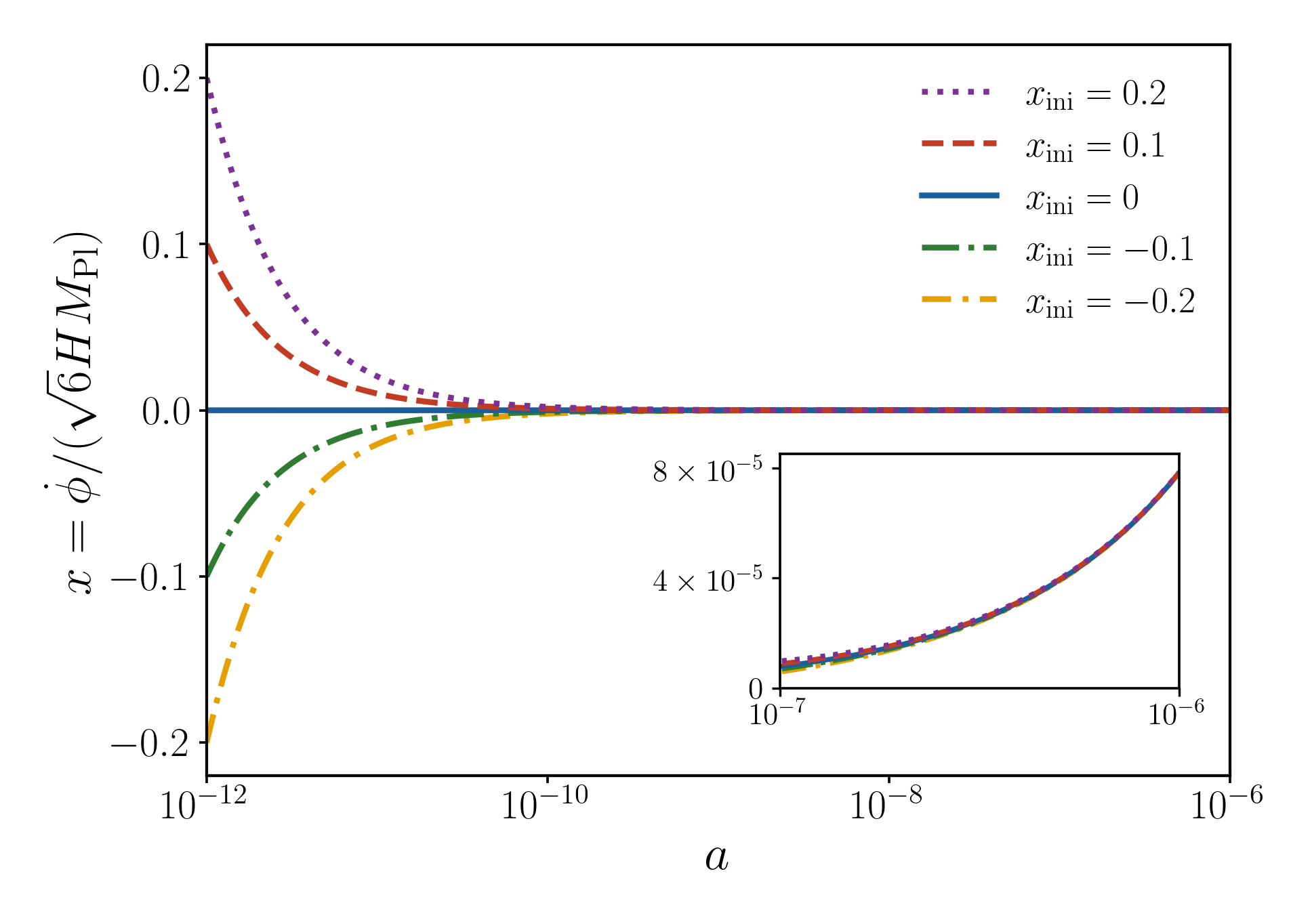}
\caption{Evolutions of   variable $x=\dot{\phi}/(\sqrt{6}H M_{\rm Pl})$ for different initial values in the CQ-EXP model. }
\label{fig1}
\end{figure}

At the very early times of the universe, the Hubble friction term  will dominate the dynamics of the scalar field due to  the large value of the Hubble parameter during this epoch.  This high value of the Hubble parameter effectively results in rapid damping of the motion of the scalar field~\cite{antusch2026guide,ramadan2024desi,caldwell2005limits}. Consequently, we can impose a nearly frozen initial condition in the radiation-dominated era, for example, $x_{\rm ini}\simeq0$ at $a_{\rm ini}\simeq10^{-12}$,  when there is no coupling between the two dark sectors, where $x\equiv\dot{\phi}/(\sqrt{6}H M_{\rm Pl})$.
In the coupled case, however, the field is not simply fixed at $x_{\rm ini} \simeq 0$. For instance, in the radiation-dominated era, we have  $H'/H\simeq-2$ and $V \leq \rho_{\phi} \ll (3M_{\rm Pl}^2H^2)$,  where a prime denotes a derivative with respect to $\ln a$. Then, Eq.~(\ref{kg_eq}) can be reduced to be
\begin{equation}\label{9}
	x'\simeq -x+\sqrt{\frac{3}{2}}\,\beta\Omega_c .
\end{equation}
Here $\Omega_c\equiv\rho_c/(3M_{\rm Pl}^2H^2)$ is the fractional  energy density of CDM.  When $\beta=0$, Eq.~(\ref{9}) has a frozen solution: $x\simeq 0$. Using $\Omega_c'\simeq\Omega_c$,  we find that Eq.~(\ref{9}) has an approximate solution 
\begin{equation}\label{10}
	x\simeq \frac{\sqrt{6}}{4}\beta\Omega_c ,
\end{equation}
which indicates that $x$ follows a frozen trajectory in the radiation-dominated era. 
 Linearizing $x$ around this trajectory,  $x \to x+\delta x$, we find that  $\delta x'\simeq-\delta x$. The negative sign indicates  a restoring behavior, suggesting that if the field deviates from its path, this deviation will decay rapidly with cosmic expansion. Therefore, the evolution of $x$ under initial conditions that differ from the trajectory  given in Eq.~(\ref{10})  will quickly return to this trajectory.  This behavior is clearly illustrated  in Fig.~\ref{fig1}, where the evolutions of  $x$ for different  initial values  are plotted.   Thus, we can use Eq.~(\ref{10}) to set  the initial value of $x$ in our analysis.

\section{Neutrino mass}
Oscillation experiments have shown that neutrinos are massive, and imposed a robust lower limit on the sum of the three known neutrino masses,  $\sum m_\nu \geq 0.059$ eV~\cite{esteban2024nufit}. Since massive neutrinos affect both the expansion history of the Universe and the growth of cosmic structure,  cosmological observations can yield strong bounds on the sum of neutrino masses.  Conversely, variation of neutrino masses also impact the constraints  placed on other cosmological parameters~\cite{planck2018cosmo,DESIDR2nu,du2025impacts,lesgourgues2006massive}. In cosmological analysis, the sum of neutrino masses is typically treated  as a positive constant, such as  $\sum m_\nu = 0.06$ eV~\cite{planck2018cosmo,DESIDR2nu}. If this assumption is relaxed while imposing the physical prior $\sum m_\nu\geq 0$, observations favor a vanishing neutrino mass sum~\cite{DESIDR2nu,du2025impacts}.  However,   recent studies have shown that when this physical prior is relaxed, allowing an effective neutrino mass parameter $\sum m_{\rm \nu, eff}$ to explore negative values, the likelihood distribution peaks in the unphysical region $\sum m_{\rm \nu, eff}<0$~\cite{DESIDR2nu,jhaveri2025negative,elbers2025negative,kibris2026negative}. In this work, we therefore consider two treatments of the neutrino sector.  Firstly, the neutrino mass sum is fixed at $\sum m_\nu=0.06\,{\rm eV}$.  Secondly, following Refs.~\cite{DESIDR2nu,elbers2025negative}, we introduce a signed effective neutrino mass  parameter, $\sum m_{\nu,\mathrm{eff}}$,  and treat it as a free parameter.

\section{Observational data}

We will constrain the cosmological models with a joint data set composed of DESI DR2 BAO measurements, DES-Dovekie SNe Ia, and Planck CMB observations.  

\begin{itemize}
\item \textbf{DESI DR2 BAO}: The BAO data are based on the first three years of DESI observations and extract distance information from multiple tracers, including bright galaxies, luminous red galaxies, emission line galaxies, quasars, and Ly$\alpha$ forest samples, thereby mapping the expansion history over a wide redshift interval~\cite{DESIDR2}.

\item \textbf{DES-Dovekie SNe Ia}: For the SNe Ia data, we consider the DES-Dovekie  compilation, which recalibrates the DES five-year supernova sample with improved photometric cross calibration~\cite{desdovekie2025,dessn5yr2024}.

 \item \textbf{Planck CMB:}  We will use the Planck temperature and polarization measurements together with lensing information, including the 2018 likelihoods, the PR4/NPIPE temperature and polarization analysis, and the PR4 lensing reconstruction~\cite{planck2018cosmo,rosenberg2022pr4camspec,carron2022pr4lensing}.

\end{itemize}

We perform the MCMC analysis with Cobaya~\cite{torrado2021cobaya} and modify CAMB to compute the CMB power spectra~\cite{amendola2000perturbations,ma1995cosmological,lewis2000camb,li2014ppf,li2023idecamb}. It is  required that  the chains  satisfy the convergence criterion $R-1<0.02$.  For cosmological model, its minimum value of  $\chi^2$  ($\chi^2_{\rm min}$ )   is obtained from the Cobaya minimizer based on the \texttt{bobyqa} algorithm.

\section{Results and discussions}
\label{sec3}

We now present the observational constraints on the CQ-EXP model from data described in above section. 
For comparison, we also analyze the $\Lambda$CDM and $w_0w_a$CDM models. The uniform prior ranges adopted for the cosmological and model parameters are summarized in Table~\ref{tab1}. We use $\Delta\chi^2$, defined the difference of $\chi^2_{\rm min} $relative to the $\Lambda$CDM model, to compare the different models. All results are summarized in Tables~\ref{tab2} and \ref{tab3}.

\begin{table}[t]
\centering
\caption{Uniform priors for parameters used in the MCMC. $\Omega_b$ and $\Omega_c$  are the present density parameters of baryon and CDM, respectively,    $\tau$ denotes the optical depth to reionization, and $A_s$ and $n_s$ are the amplitude and spectral index of primordial curvature  perturbations, respectively.} 
\label{tab1}
\renewcommand{\arraystretch}{1.25}

\newcommand{\wA}{2.10cm}
\newcommand{\wB}{2.40cm}
\newcommand{\wC}{1.0cm}
\newcommand{\wD}{2.5cm}

\begin{tabular}{l c|l c}
\hline\hline
\multicolumn{2}{c|}{Cosmological Parameters} & \multicolumn{2}{c}{Model Parameters} \\
\hline
\makebox[\wA][l]{$\Omega_b h^2$} & \makebox[\wB][c]{$U[0.005,\,0.1]$}
& \makebox[\wC][c]{$w_0$} & \makebox[\wD][c]{$U[-3,\,1]$} \\

\makebox[\wA][l]{$\Omega_c h^2$} & \makebox[\wB][c]{$U[0.001,\,0.99]$}
& \makebox[\wC][c]{$w_a$} & \makebox[\wD][c]{$U[-3,\,2]$} \\

\makebox[\wA][l]{$H_0$} & \makebox[\wB][c]{$U[20,\,100]$}
& \makebox[\wC][c]{$\alpha$} & \makebox[\wD][c]{$U[0,\,2]$} \\

\makebox[\wA][l]{$\ln(10^{10} A_s)$} & \makebox[\wB][c]{$U[1.61,\,3.91]$}
& \makebox[\wC][c]{$\beta$} & \makebox[\wD][c]{$U[-0.5,\,0.5]$} \\

\makebox[\wA][l]{$n_s$} & \makebox[\wB][c]{$U[0.8,\,1.2]$}
& & \\

\makebox[\wA][l]{$\tau$} & \makebox[\wB][c]{$U[0.01,\,0.8]$}
& & \\

\makebox[\wA][l]{$\sum m_{\nu,\mathrm{eff}}$} & \makebox[\wB][c]{$U[-0.5,\,0.5]$}
& & \\
\hline\hline
\end{tabular}
\end{table}

For the $\Lambda$CDM model, with the sum of neutrino masses fixed at $\sum m_\nu=0.06\,{\rm eV}$, our analysis yields 
\begin{eqnarray}\label{eq12}
\Omega_m=0.3037\pm 0.0035, \quad H_0=68.09\pm0.27 ~{\rm km/s/Mpc}, \quad S_8=0.8107\pm 0.0078.
\end{eqnarray}
Here, $\Omega_m$ represents the present matter density parameter, encompassing baryon, CDM, and massive neutrino. The value of the Hubble constant $H_0$ is significantly lower than $H_0 = 73.04 \pm 1.04$~km/s/Mpc obtained from nearby SNe Ia and Cepheid variables~\cite{riess2022shoes}. Additionally, the derived value of $S_8$ is substantially larger than those from weak lensing surveys~\cite{DESY3S8,kids1000}. Here, $S_8\equiv\sigma_8 \sqrt{\frac{\Omega_{m}}{0.3}}$ with  $\sigma_8$ representing the root mean square fluctuation of matter density in spheres of radius  $8h^{-1}$ Mpc and  $h\equiv H_0/(100 ~\rm km/s/Mpc)$. These discrepancies indicate that the $\Lambda$CDM model is suffering both the $H_0$ tension\cite{verde2019tensions,knox2020hubble,divalentino2021h0review,schoneberg2022h0olympics,perivolaropoulos2022challenges} and the growth tension~\cite{perivolaropoulos2022challenges,divalentino2021s8,abdalla2022tensions}. 

When the effective neutrino mass parameter is treated as a free parameter, we find  from Table~\ref{tab3} that
\begin{eqnarray}
\sum m_{\nu, \rm eff}=-0.068^{+0.049}_{-0.061} ~{\rm eV}, 
\end{eqnarray}
indicating that observations favor a negative mass parameter at more than $1\sigma$ CL, which is consistent with what have been found in \cite{DESIDR2nu,jhaveri2025negative,elbers2025negative,kibris2026negative}. The value of $\Omega_m$ is consistent with $\Omega_m=0.3037\pm 0.0035$ obtained  when the neutrino mass sum is fixed. As a result of allowing the neutrino mass to vary, the values of $H_0$ and $S_8$ increase slightly. Thus, this additional free parameter  slightly alleviates the $H_0$ tension, while exacerbating the growth tension. Furthermore, we find that $\Delta\chi^2=-6.04$, suggesting that a varying neutrino mass is favored by the observations. 

In the framework of the $w_0w_a$CDM model with a fixed $\sum m_\nu$, we achieve   $\Omega_m=0.3132\pm 0.0053$ and $S_8=0.8244\pm 0.0087$, both of  which are slightly larger than those derived from the $\Lambda$CDM model. In contrast, the value of $H_0=67.31\pm 0.54$~km/s/Mpc is lower than the one presented in Eq.~(\ref{eq12}). Thus, the tensions related to both $H_0$ and $S_8$ become more pronounced in the $w_0w_a$CDM model. The model parameters $w_0$ and $w_a $ are constrained to be 
\begin{eqnarray}
w_0= -0.811\pm 0.055,\quad w_a=-0.69^{+0.23}_{-0.20}, 
\end{eqnarray}
showing that a dynamical dark energy is favored, along with a present dark energy EoS  that exceeds $-1$. This conclusion is further supported by the $\Delta \chi^2$ value of $-14.48$, which suggests that a dynamical dark energy improve the fit to the observational data. When the neutrino mass is allowed to vary, the results show minimal change. A zero effective  neutrino mass parameter  remains a possibility, as indicated by $\sum m_{\nu, \rm eff}=-0.036\pm0.088$ eV. Additionally, since the value of $\Delta \chi^2$ only has a negligible decrease when $\sum m_{\nu, \rm eff}$ is treated as a free parameter, the observations seems to disfavor the scenario of a varying total neutrino mass parameter within the $w_0w_a$CDM model. This contrasts with the results obtained in the $\Lambda$CDM model.

\begin{table*}[t]
\centering
\caption{Summary of the marginalized results  for a fixed $\sum m_\nu=0.06\,{\rm eV}$. }
\label{tab2}
\small
\setlength{\tabcolsep}{5pt}
\begin{tabular}{lcccc}
\hline\hline
Parameter & $\Lambda$CDM & $w_0w_a$CDM &
\begin{tabular}{c}CQ-EXP\\$\beta>0$\end{tabular} &
\begin{tabular}{c}CQ-EXP\\$\beta<0$\end{tabular} \\
\hline
$w_0$ & $-$ & $-0.811\pm 0.055$ & $-$ & $-$ \\
$w_a$ & $-$ & $-0.69^{+0.23}_{-0.20}$ & $-$ & $-$ \\
$\alpha$ & $-$ & $-$ & $0.62^{+0.20}_{-0.12}$ & $0.81^{+0.21}_{-0.12}$ \\
$\beta$ & $-$ & $-$ & $0.047^{+0.014}_{-0.008}$ & $-0.045^{+0.007}_{-0.012}$ \\
$H_0$ & $68.09\pm 0.27$ & $67.31\pm 0.54$ & $67.75\pm 0.57$ & $67.66 \pm {0.57}$ \\
\hline\hline
$\Omega_m$ & $0.3037\pm 0.0035$ & $0.3132\pm 0.0053$ & $0.3012\pm 0.0055$ & $0.3079 \pm {0.0058}$ \\
$S_8$ & $0.8107\pm 0.0078$ & $0.8244\pm 0.0087$ & $0.8189\pm 0.0086$ & $0.8254\pm 0.0097$ \\
\hline\hline
$\Delta \chi^2$ & $0$ & $-14.48$ & $-10.21$ & $-12.62$ \\
\hline\hline
\end{tabular}
\end{table*}

\begin{table*}[t]
\centering
\caption{Summary of the marginalized results when the effective neutrino mass parameter $\sum m_{\nu,\mathrm{eff}}$ is treated as a free parameter.}
\label{tab3}
\small
\setlength{\tabcolsep}{5pt}
\begin{tabular}{lcccc}
\hline\hline
Parameter & $\Lambda$CDM & $w_0w_a$CDM &
\begin{tabular}{c}CQ-EXP\\$\beta>0$\end{tabular} &
\begin{tabular}{c}CQ-EXP\\$\beta<0$\end{tabular} \\
\hline
$w_0$ & $-$ & $-0.844\pm0.061$ & $-$ & $-$ \\
$w_a$ & $-$ & $-0.48^{+0.29}_{-0.26}$ & $-$ & $-$ \\
$\alpha$ & $-$ & $-$ & $0.69^{+0.20}_{-0.11}$ & $0.81^{+0.18}_{-0.13}$ \\
$\beta$ & $-$ & $-$ & $0.025^{+0.019}_{-0.025}$ & $-0.028^{+0.028}_{-0.018}$ \\
$\sum m_{\nu,\mathrm{eff}}$ & $-0.068^{+0.049}_{-0.061}$ & $-0.036\pm0.088$ & $-0.073^{+0.057}_{-0.097}$ & $-0.052^{+0.068}_{-0.10}$ \\
$H_0$ & $68.64\pm0.38$ & $67.44\pm0.56$ & $67.70\pm0.57$ & $67.62\pm0.56$ \\
\hline\hline
$\Omega_m$ & $0.2982\pm0.0043$ & $0.3099\pm0.0061$ & $0.3028\pm0.0055$ & $0.3070\pm0.0056$ \\
$S_8$ & $0.828\pm0.010$ & $0.831\pm0.011$ & $0.830\pm0.011$ & $0.832\pm0.011$ \\
\hline\hline
$\Delta \chi^2$ & $-6.04$ & $-14.49$ & $-12.90$ & $-12.80$ \\
\hline\hline
\end{tabular}
\end{table*}

Now, we study the constraints on the CQ-EXP model. Since different values of $\beta$  result in  different dynamics of the scalar field, we separate our analysis into two cases:   $\beta>0$  and   $\beta<0$.

\subsection{\texorpdfstring{$\beta>0$}{beta > 0}}

We first consider the case of $\sum m_\nu=0.06\,{\rm eV}$. From  Table~\ref{tab2}, we find that the constraints on $\Omega_m$, $H_0$ and $S_8$ are in good agreement with  those obtained in the $\Lambda$CDM model. The parameters $\alpha$ and $\beta$ are constrained  to 
\begin{equation} \label{14}
\alpha=0.62^{+0.20}_{-0.12},\qquad
\beta=0.047^{+0.014}_{-0.008},
\end{equation}
at the $1\sigma$ CL. Thus, the coupling between quintessence and CDM is favored by the observations since $\beta$ deviates from zero at  more than $3\sigma$ CL. 

\begin{figure}[t]
\centering
\includegraphics[width=0.48\textwidth]{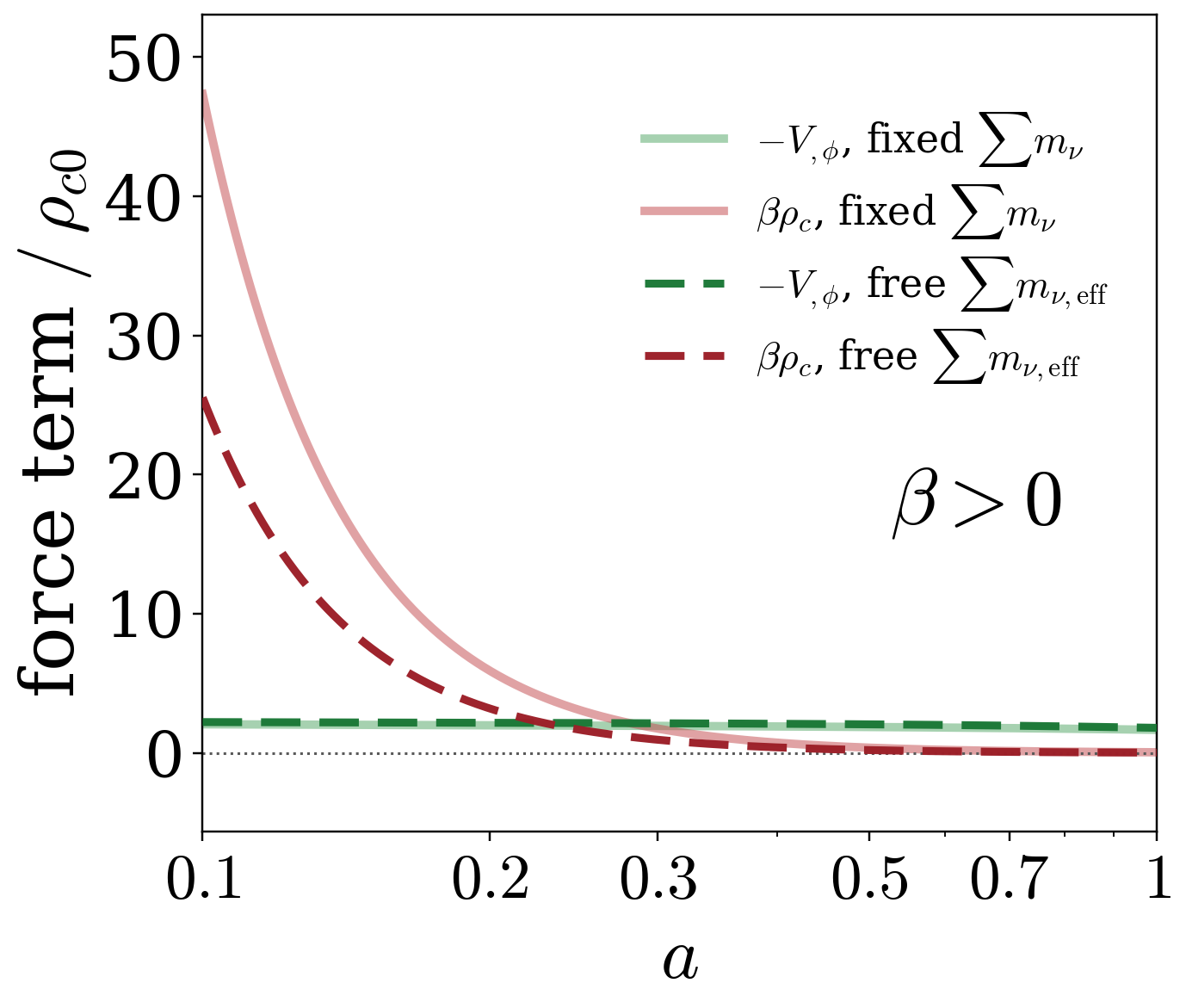}
\caption{Evolutions of the  force from the potential and the effective force from the interaction between the two dark sectors for the $\beta>0$ case.}
\label{fig2}
\end{figure}

\begin{figure*}[t]
\centering
\begin{minipage}{0.48\textwidth}
\centering
\includegraphics[width=\linewidth]{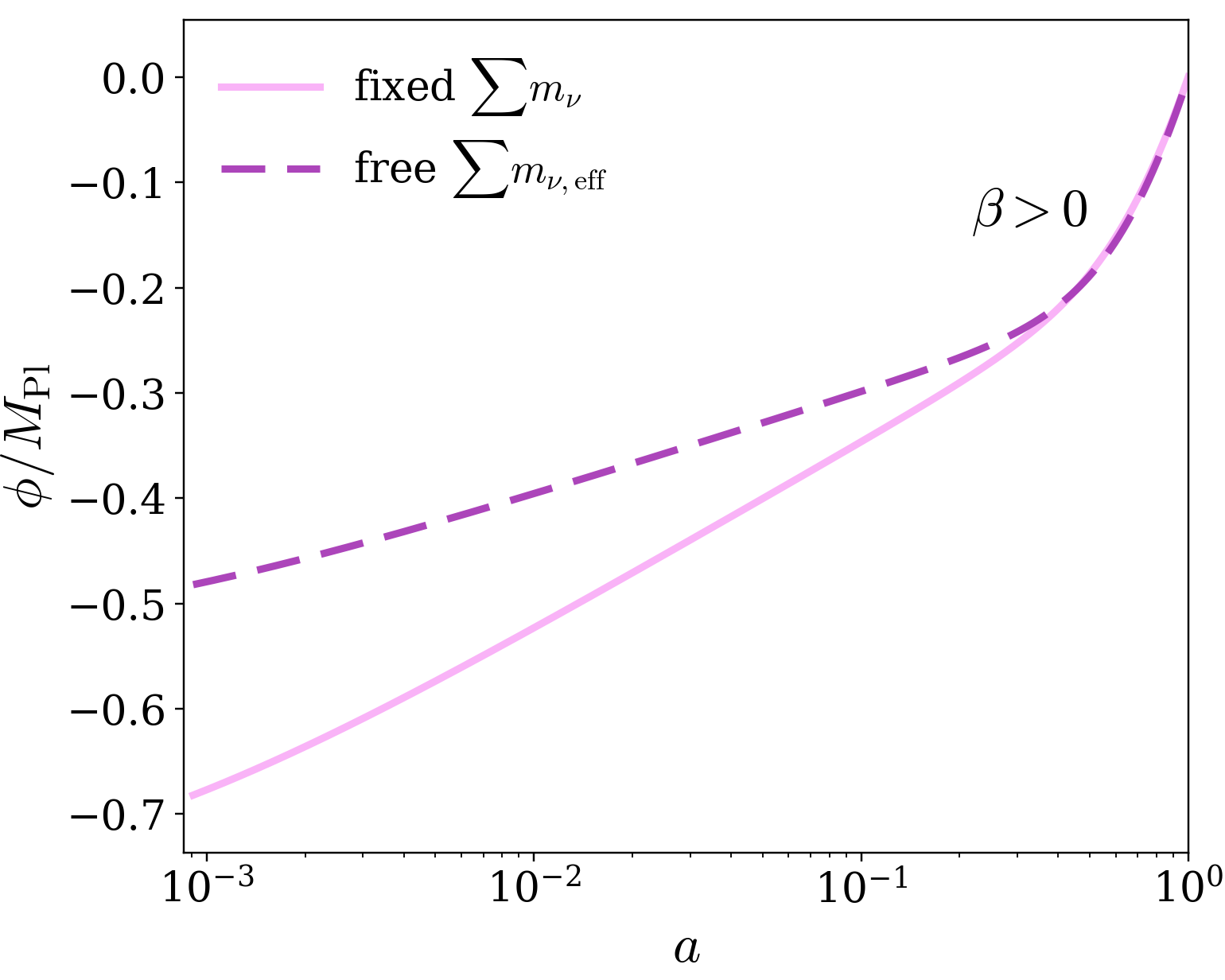}
\end{minipage}\hfill
\begin{minipage}{0.48\textwidth}
\centering
\includegraphics[width=\linewidth]{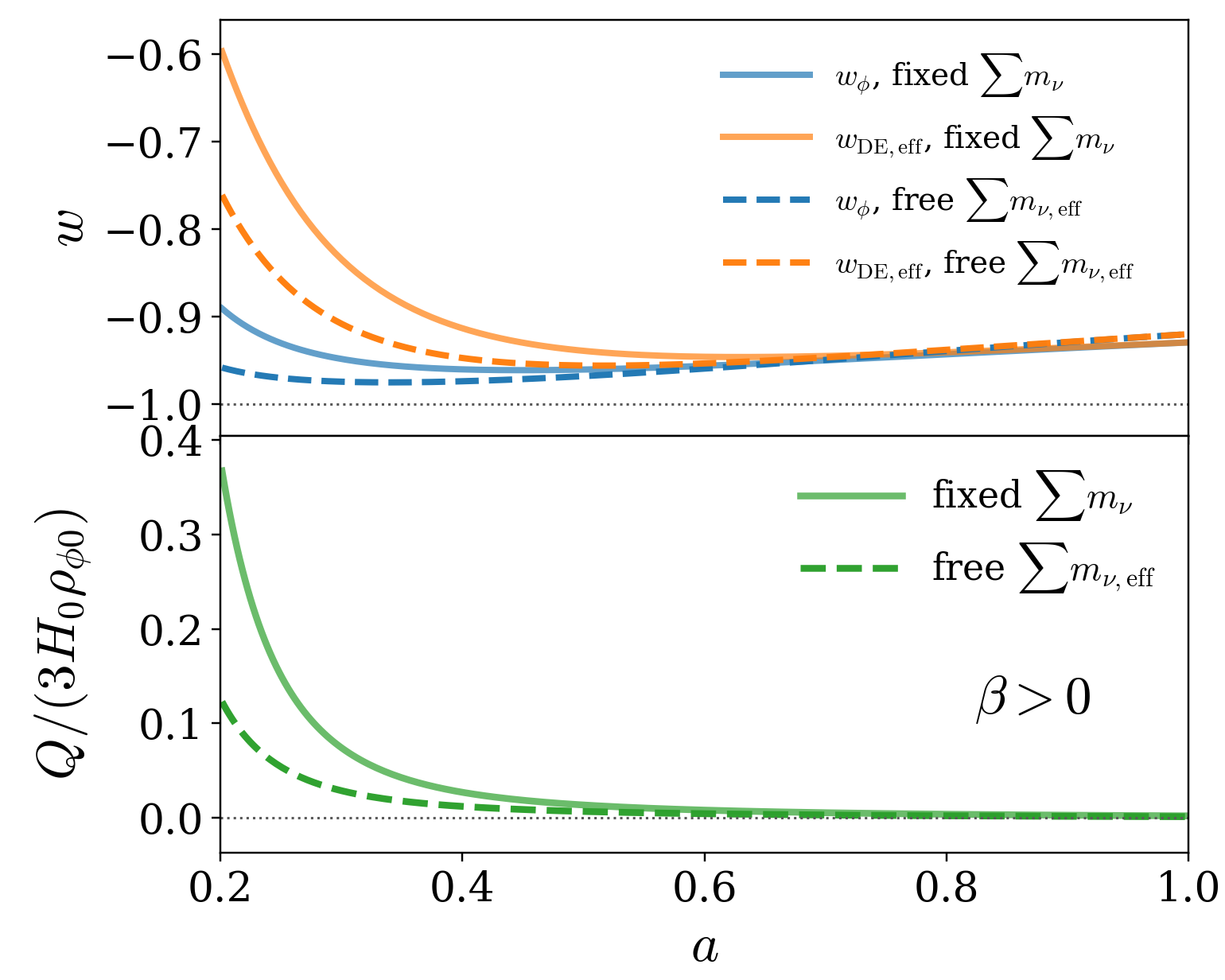}
\end{minipage}
\caption{Evolutions of $\phi$, $w_\phi$, $w_{{\rm DE},{\rm eff}}$ and $Q$ for the case of $\beta>0$.}
\label{fig3}
\end{figure*}

When $\alpha>0$ and $\beta>0$, Eq.~(\ref{kg_eq}) demonstrates that both the conservative force derived from the exponential potential, $-V_{,\phi}$, and the effective force induced by the interaction, $\beta\rho_c/M_{\rm Pl}$,  are positive and act in the same direction within field space. This is also  illustrated in Fig.~\ref{fig2}, which shows the evolution of  both the conservative force and  the effective force as the universe expands.  At  early times, the effective force dominates  because it is proportional to the CDM energy density and this matter dominates the cosmic evolution in the matter-dominated era.  However, it  decreases rapidly with cosmic expansion, while the potential force varies much more slowly,  ultimately becoming the dominant driving force at late times.

This characteristic of the conservative force and  the effective force means that  the scalar field rolls monotonically toward larger values of $\phi$, as shown in the left panel of Fig.~\ref{fig3}, with $\dot{\phi}$ remaining positive throughout the evolution.  A positive $\dot{\phi}$ ensures that the value of $Q$ is always positive, even though it decreases with cosmic expansion, as clearly illustrated  in the right panel of Fig.~\ref{fig3}. Consequently,  energy  is continuously transferred  from CDM to the scalar field.  As a result, both the EoS of the scalar field, $w_\phi$,  and the effective EoS of dark energy, $w_{{\rm DE},{\rm eff}}$  remain greater than $-1$,  even as they evolve with cosmic expansion, as represented in  the right panel of Fig.~\ref{fig3}. Thus, the coupling between the scalar field and CDM with a positive $\beta$ does not allow for a crossing of the $-1$ line for the effective EoS of quintessence dark energy. Nevertheless, the value of $\Delta\chi^2$ reaches $-10.21$,   indicating that the observations favor the CQ-EXP model with a positive $\beta$ over the $\Lambda$CDM model.  

After treating  the effective neutrino mass parameter as a free parameter, we summarize the results in table~\ref{tab3} and find that 
\begin{eqnarray}
\sum m_{\nu,\mathrm{eff}}&=-0.073^{+0.057}_{-0.097},
\end{eqnarray}
which deviates from zero at more than $1\sigma$ and  is comparable  with the result obtained in the $\Lambda$CDM model. This flexibility in neutrino mass has negligible effects on the constraints for $\Omega_m$, $H_0$ and $S_8$,  and thus does not  alleviate  either the $H_0$ tension or the growth tension. However,  this additional freedom does affect   the allowed regions of $\alpha$ and $\beta$, which become 
\begin{equation}
\begin{aligned}
\alpha&=0.69^{+0.20}_{-0.11},\qquad
\beta&=0.025^{+0.019}_{-0.025}.\end{aligned}
\end{equation}
Comparing these results  with Eq.~(\ref{14})  obtained under the fixed neutrino mass assumption reveals that the preference for $\beta>0$ is reduced from more than $3\sigma$ to within $1\sigma$.  These findings indicate a strong degeneracy between the effective neutrino mass and the coupling between quintessence and CDM~\cite{pettorino2012constraints}. Since a smaller $\beta$ is obtained,  the effective force, proportional to $\beta\rho_c/M_{\rm Pl}$, is therefore suppressed, allowing the conservative force from the potential to dominate earlier. This scenario is clearly illustrated  in Fig.~\ref{fig2}. So,   the scalar field evolves more slowly during the early times, as shown in right panel of Fig.~\ref{fig3}.  The reduced field velocity lowers the kinetic contribution, shifting both $w_\phi$ and $w_{{\rm DE},{\rm eff}}$ closer to $-1$ in the right panel of Fig.~\ref{fig3}, while the energy transfer rate $Q$ is also suppressed.   
Additionally, we find that the value of $\Delta\chi^2$ is $\Delta\chi^2=-12.90$,  which is less apparent than $\Delta\chi^2=-10.21$ obtained with the fixed neutrino mass sum case.  Thus, in the CQ-EXP model with $\beta>0$, the observations seem to support a variation in neutrino mass,  consistent with findings   in the   $\Lambda$CDM model,  but differing from those in the $w_0w_a$CDM model.

\subsection{\texorpdfstring{$\beta<0$}{beta < 0}}

Table~\ref{tab2}, where the value of $\sum m_\nu$ is  fixed,  clearly shows that the  constraints on $\Omega_m$, $H_0$ and $S_8$ are consistent  with those obtained in the case of $\beta>0$. The allowed values of $\alpha$ and $\beta$ are given by 
\begin{equation} 
\alpha=0.81^{+0.21}_{-0.12},\qquad
\beta=-0.045^{+0.007}_{-0.012}.
\end{equation}
 Notably,  the value of $\alpha$ is larger than that obtained in the $\beta>0$ case, and  the coupling parameter $\beta$ is separated from zero by more than $4\sigma$. 

From table \ref{tab3}, which shows the results in the case of   $\sum m_\nu$  treated as free parameter, we find that 
\begin{equation}
\sum m_{\nu,\mathrm{eff}}=-0.052^{+0.068}_{-0.100}
\end{equation}
which is less than zero but $\sum m_{\nu,\mathrm{eff}}=0$ is allowed at the $1\sigma$ CL. This freedom does not affect significantly the constraints on $\Omega_m$, $H_0$,  $S_8$ and $\alpha$, which aligns well with those obtained in the case of $\sum m_\nu=0.06$~eV. They are also compatible with what are achieved in the  $\beta>0$ case. While, the freedom of the neutrino mass has an important impact on the result of  $\beta$,  which is 
\begin{equation}
\beta=-0.028^{+0.028}_{-0.018}.
\end{equation}
This value indicates that the non-coupling case is possible at the $1\sigma$ CL.

\begin{figure}[t]
\centering
\includegraphics[width=0.48\textwidth]{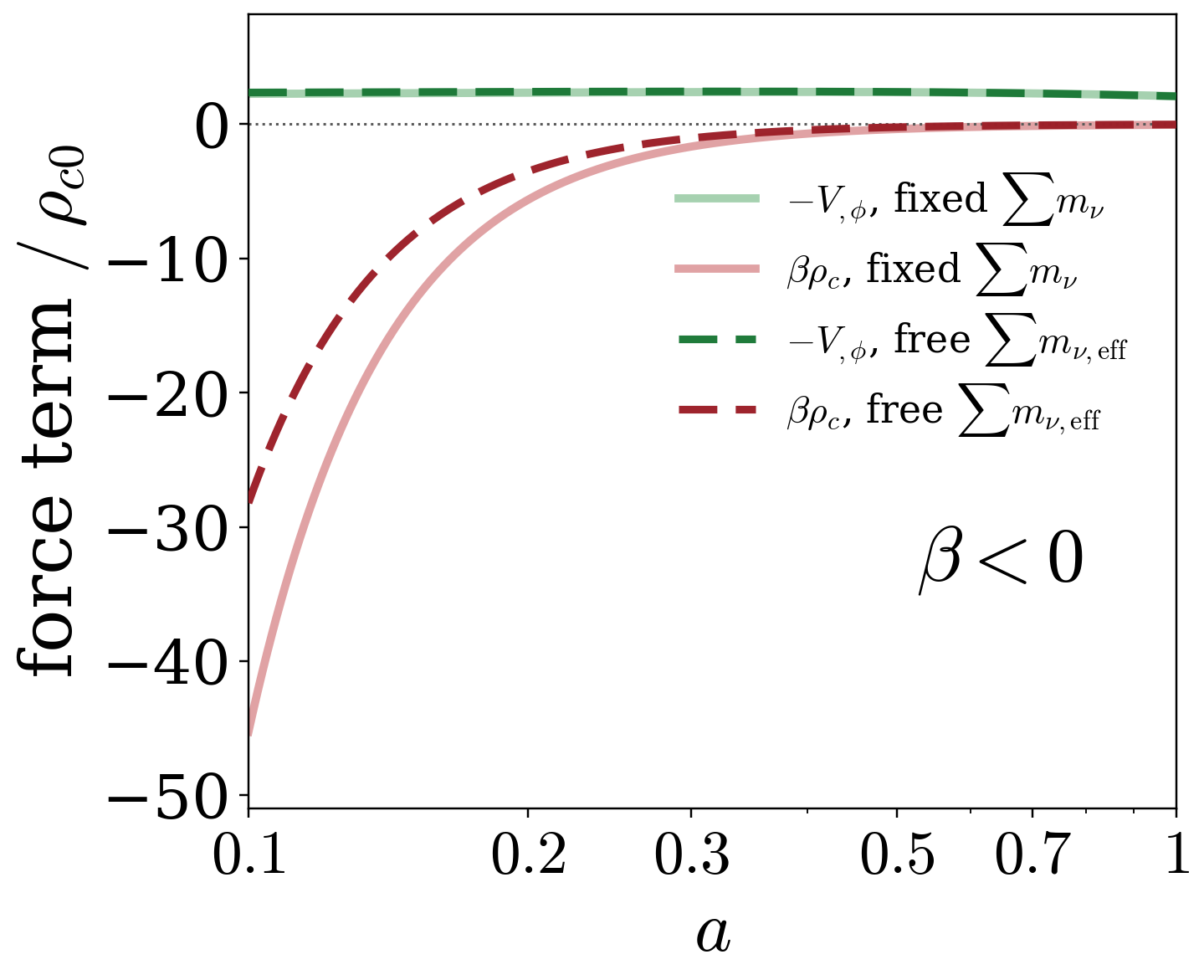}
\caption{Evolutions of the  force from the potential and the effective force from the interaction between the two dark sectors for the case of $\beta<0$.}
\label{fig4}
\end{figure}

\begin{figure*}[t]
\centering
\begin{minipage}{0.48\textwidth}
\centering
\includegraphics[width=\linewidth]{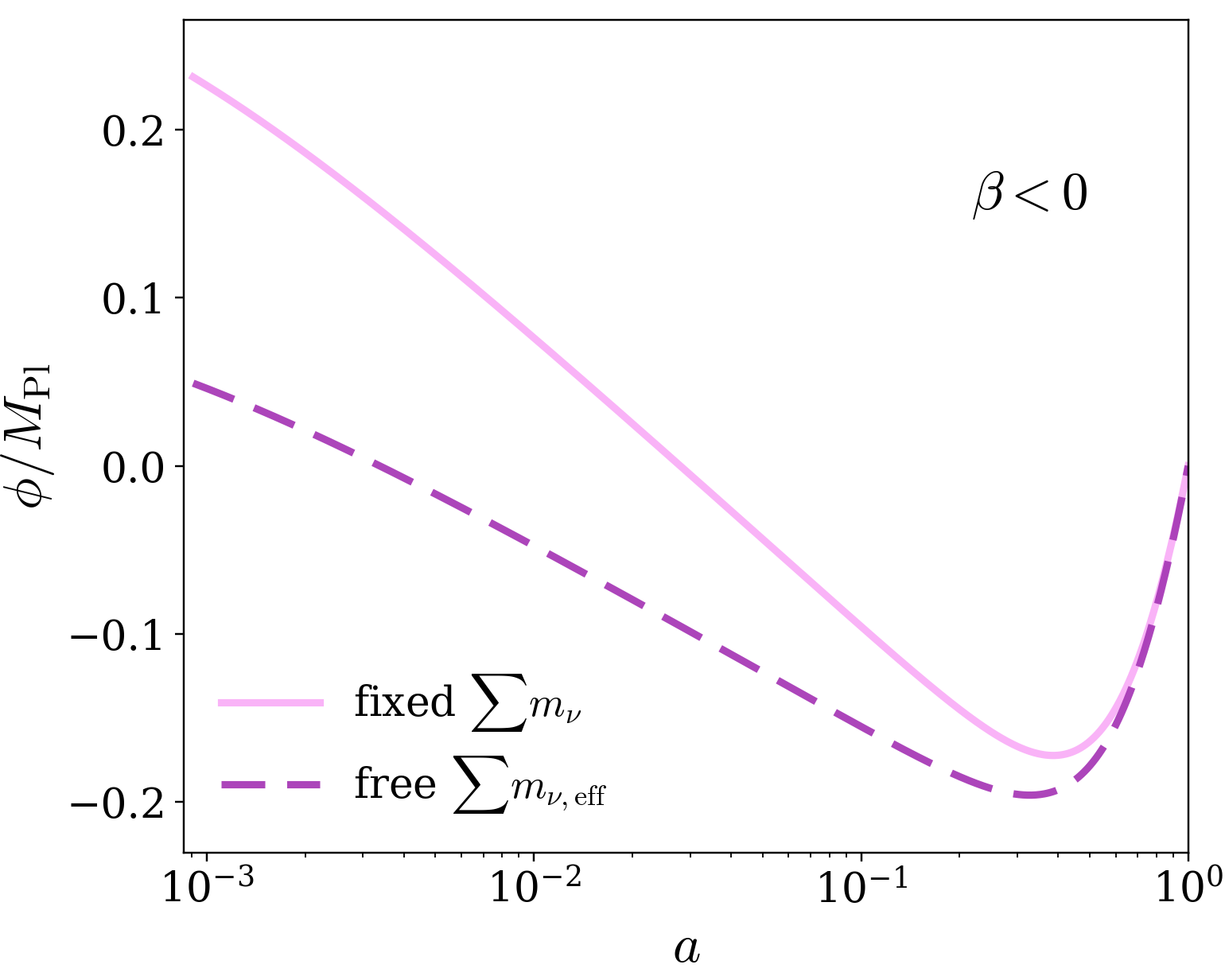}
\end{minipage}\hfill
\begin{minipage}{0.48\textwidth}
\centering
\includegraphics[width=\linewidth]{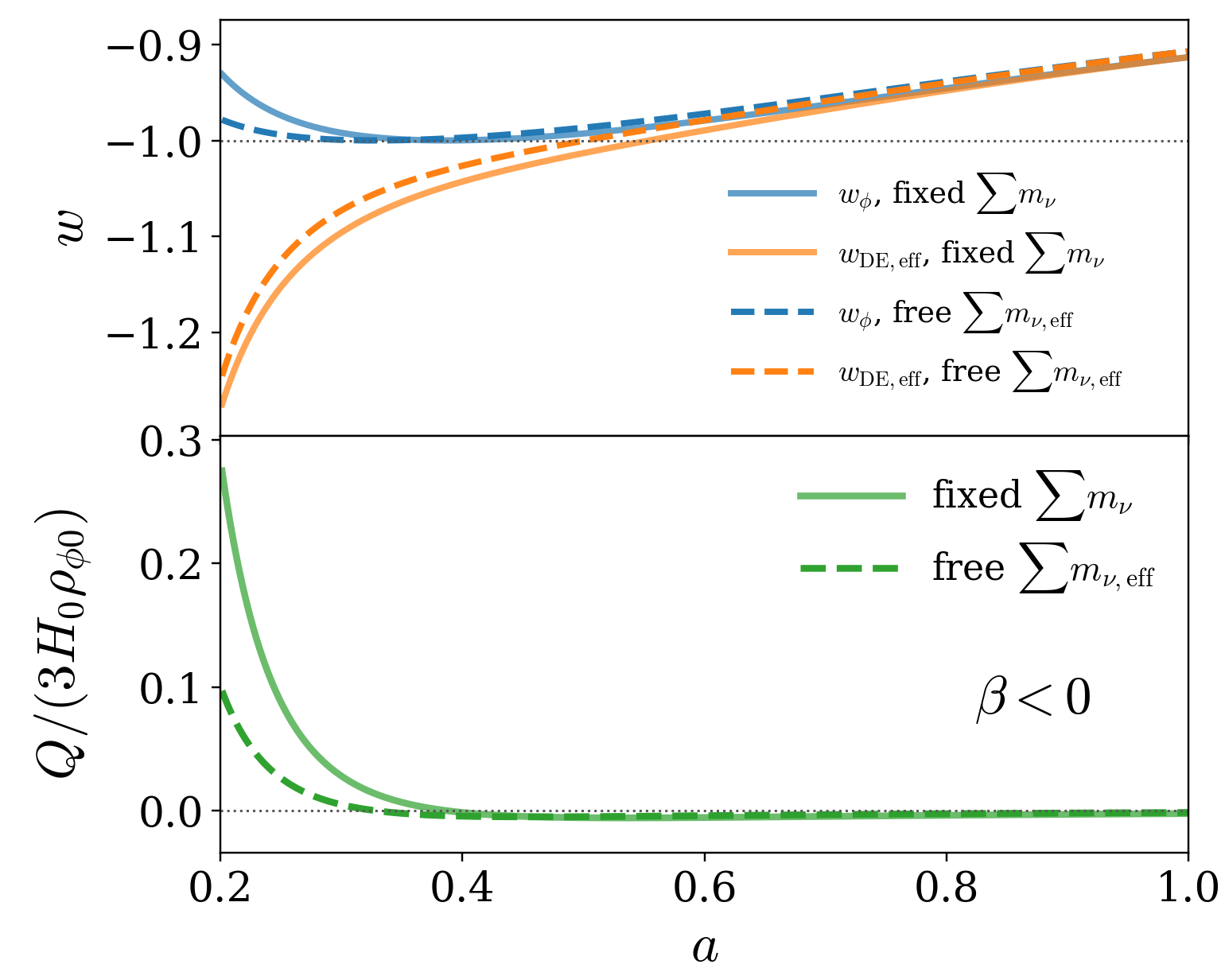}
\end{minipage}
\caption{Evolutions of $\phi$, $w_\phi$, $w_{{\rm DE},{\rm eff}}$, and $Q$ for the $\beta<0$ case.}
\label{fig5}
\end{figure*}

In this case,  the effective force from the interaction between quintessence and CDM and the potential force now act in opposite directions. This effective force dominates  the evolution of scalar field in the early times due to the large value of $\rho_c$ and drives the scalar field rolling up along its potential. With the cosmic expansion, the absolute value of the effective force decreases rapidly while the potential force varies slowly. This scenario has shown in Fig.~\ref{fig4}. Thus, the potential force will dominate the evolution of scalar field and leads  the scalar field to roll down along the potential. Therefore, the evolution of the scalar field bounces with cosmic expansion, as shown in the left panel of Fig.~\ref{fig5}.   This bounce behavior causes $Q$ to evolve from positive to negative with cosmic expansion, which indicates that at the early times the energy transfers from  CDM to the scalar field, and from quintessence to CDM at the late times. The sign-changing interaction between quintessence and CDM allows $w_{\rm DE, eff}$ to cross the $-1$ line although $w_\phi$ remains larger than $-1$. This scenario can be seen in the right panel of Fig.~\ref{fig5}.

The values of $\Delta\chi^2$ are $\Delta\chi^2=-12.62$ for the fixed $\sum m_\nu$ case and  $\Delta\chi^2=-12.80$ for the  free $\sum m_{\nu,\mathrm{eff}}$ case.   The small  differences between these values  indicate that allowing the effective neutrino mass parameter to vary does not significantly improve the fit of this $\beta<0$ branch.

\section{Conclusions}%
\label{sec4}

In this work, we have studied the CQ-EXP model,   in which quintessence with an exponential potential is coupled to CDM through an exponential coupling function, $e^{-\beta \phi/M_{\rm Pl}}$. Although the coupling parameter $\beta$ is usually assumed to be positive~\cite{amendola2000coupled,pettorino2008cqe,baldi2012codecs,beltran2026ceq,liu2024emtransfer}, we extend  the analysis to include the branch with $\beta<0$. We have shown that, in the radiation-dominated era, the scalar field follows a frozen trajectory, as described in Eq.~(\ref{10}), which is then used as the initial condition for the subsequent cosmological evolution.

We have  tested  the CQ-EXP model  using  Planck CMB, DESI DR2 BAO, and DES-Dovekie SNe Ia observations. We find that  the constraints on cosmological parameters, such as $\Omega_m$, $H_0$ and $S_8$, are consistent with those obtained in the $\Lambda$CDM and $w_0w_a$CDM models. Therefore, both the $H_0$ tension  and the growth tension  persist  in the CQ-EXP model.   When the sum of neutrino mass is fixed at $\sum m_\nu=0.06\,{\rm eV}$, the data favor  a coupling between quintessence dark energy and CDM, as indicated by the fact that the coupling parameter $\beta$ deviates from zero at more than $3\sigma$. For $\beta>0$, the effective EoS of dark energy evolves with cosmic expansion but  does not cross the phantom divide line. In contrast, for $\beta<0$, $w_{\rm DE, eff}$ can evolve from below $-1$ to above $-1$. This behavior occurs  because, when $\beta<0$ the effective force resulting from the coupling between dark energy and CDM acts in the  direction opposite  to the force generated by the scalar-field potential. Consequently,  the scalar can initially  roll up its potential before subsequently rolling down, causing its rolling speed to change sign.   This sign change leads to a sign-changing energy transfer between the two dark sectors and enables $w_{\rm DE, eff}$ to cross the $-1$ line. Additionally, the value of $\Delta\chi^2$ for the CQ-EXP model with $\beta<0$ is  smaller than that for the model with $\beta>0$. Therefore,  for a fixed neutrino mass sum, current observations favor a dynamical dark-energy scenario in which the effective EoS crosses the $-1$ line. We also find that the CQ-EXP model with $\beta<0$ is statistically indistinguishable from dark energy described by the CPL parametrization, with only a very small difference in $\chi^2_{\rm min}$.

We have further analyzed the scenario in which   the effective neutrino mass sum parameter  $\sum m_{\nu,\mathrm{eff}}$ is treated as a free parameter. Our results indicate that a preference for  negative values of $\sum m_{\nu,\mathrm{eff}}$. In the CQ-EXP model with $\beta>0$,  the  $\sum m_{\nu,\mathrm{eff}}$ deviates  from zero at more than $1\sigma$ CL, similar to the result  obtained in the $\Lambda$CDM model. By contrast, in the CQ-EXP model with $\beta<0$ and in the $w_0w_a$CDM model, $\sum m_{\nu,\mathrm{eff}}=0$  remains  allowed within $1\sigma$. Allowing  $\sum m_{\nu,\mathrm{eff}}$  to vary   weakens the preference for a coupling between quintessence and CDM, since  $\beta=0$ becomes marginally allowed at the $1\sigma$ level. Interestingly, this additional freedom significantly increases  the absolute value of  $\Delta\chi^2$ for the CQ-EXP model with $\beta>0$, while its effect on  the $\Delta\chi^2$ for the CQ-EXP model with $\beta<0$ and the $w_0w_a$CDM model is negligible. As a result, the $\Delta\chi^2$ values for the $\beta>0$ and $\beta<0$ CQ-EXP models are nearly identical,  both being only about 1.6  larger than that of the $w_0w_a$CDM model. This small difference suggests that   the CQ-EXP model cannot be statistically distinguished from the $w_0w_a$CDM model with the current data.

In conclusion,  when $\sum m_\nu=0.06$ eV is fixed,  observations favor the CQ-EXP model with $\beta<0$  over that with $\beta>0$,  because the former allows the effective EoS of dark energy to cross the $-1$ line. When $\sum m_{\nu, \rm eff}$ is treated as a free parameter, however, the $\beta>0$ and $\beta<0$ branches of the CQ-EXP model receive comparable observational support and are statistically indistinguishable from the $w_0w_a$CDM model. 
Thus,   if negative values of $\sum m_{\nu,\mathrm{eff}}$ are allowed, a dynamical dark-energy model without phantom-divide crossing in its effective EoS can also provide a successful explanation of the latest observations.

\begin{acknowledgments}
This work was supported in part by the NSFC under Grant  Nos. 12275080 and 12075084,  and the Innovative Research Group of Hunan Province under Grant No.~2024JJ1006.

\end{acknowledgments}

\end{document}